\documentclass[a4paper,aps,prl,preprint,showpacs,superscriptaddress,floatfix]{revtex4} 
\usepackage[T1]{fontenc}
\usepackage[english]{babel}
\usepackage[stable]{footmisc}

\usepackage{pslatex}

\usepackage{amsmath}
\usepackage{array}
\usepackage{subfigure}
\usepackage{longtable}

\usepackage{graphicx}
\usepackage{natbib}

\begin{document}

\title{Telechelic star polymers as self-assembling units\\ from the molecular to the macroscopic scale}

\author{Barbara Capone}
\affiliation{Faculty of Physics, University of Vienna, Boltzmanngasse 5, A-1090 Vienna, Austria}
\author{Ivan Coluzza}
\affiliation{Faculty of Physics, University of Vienna, Boltzmanngasse 5, A-1090 Vienna, Austria}
\author{Federica Lo Verso}
\affiliation{Institut f\"ur Physik, Johannes-Gutenberg-Universit\"at Mainz, D-55099 Mainz, Germany}
\author{Christos N.\ Likos}
\affiliation{Faculty of Physics, University of Vienna, Boltzmanngasse 5, A-1090 Vienna, Austria}
\author{Ronald Blaak}
\affiliation{Faculty of Physics, University of Vienna, Boltzmanngasse 5, A-1090 Vienna, Austria}

\pacs{82.70.-y, 81.16.Fg, 05.10.-a, 83.80.Uv}

\date{\today}

\begin{abstract} 
By means of multiscale molecular simulations, we show that telechelic-star polymers  are a  simple, robust  and  tunable system, which  hierarchically self-assembles first into soft-patchy particles and then into targeted crystalline structures. 
The self-aggregating patchy behavior can be fully controlled by  the number of arms per star and by the fraction of attractive monomeric units at the free ends  of the arms. Such self-assembled soft-patchy particles  while forming, upon augmenting density,  gel-like percolating networks and  stable ordered structures,  preserve  properties as  particle size, number and  arrangement of patches per particle.
In particular, we  demonstrate that the flexibility inherent in the soft-patchy particles 
brings forward a novel mechanism that leads to the stabilisation of diamond and simple cubic crystals
over a wide range of densities, and for molecular sizes ranging from about 10 nm up to the micrometer scale.
\end{abstract}

\maketitle

The current design of materials in the nanometer scale with its vast applications for, e.g., molecular electronics, therapeutic vectors and diagnostic imaging agent carriers, or photonics,
focuses on the fabrication of a variety of organic and inorganic building blocks with different sizes and shapes, which are chemically decorated with discrete and specific interaction sites.
Over the past years, functionalised particles gained an important role as  building blocks for various materials \cite{Glotzer07, Daan, Torquato, romano:184501, Panagiotopoulos}, rising the need for fabrication techniques that are both simple enough, and, at the same time, allow for a large-scale and reliable production of these particles.
State of the art methods for the synthesis of such particles include lithography~\cite{Velegol1,Velegol2}, microfluidics~\cite{Nie} or glancing angle deposition~\cite{Kretz_2008, Kretz_2009}. However, shape-specific particle fabrication becomes more challenging when the particle has to be decorated by a large number of 
patches or if the shape and location of the patches has to be tuned and controlled with extremely high precision.

In this work, we show that di-block copolymer stars are a macromolecular
system that allows for an extremely well-controlled production of novel particles with a potentially unlimited number of patches. 
Diblock copolymer stars are star-polymers whose arms are characterized by dual physical or chemical functionalities. 
In particular, each arm is an amphiphlic $AB$-diblock with a solvophilic, athermal part $A$ that is grafted in the center of the star, followed by  a solvophobic, thermosensitive part $B$. Accordingly, we
adopt the terminology {\it telechelic star polymers} (TSP) throughout. 
The solvophilic nature of the interior monomers results in a preferential maximization of the exposure to the solvent, whereas the solvophobic tail monomers tend to minimize their interaction with the same.
By employing a coarse-grained approach, we surpass the molecular size limits of previous computational and theoretical approaches \cite{PRL-FG,faraday} demonstrating  how, for extremely large and very dense systems  (up to $10^7$ monomers), by modifying  the  chemical composition of the polymer chains and the number of arms, it is possible to obtain stable soft-patchy particles that  survive in their self-assembled patchy state up to high concentrations. In contrast to the conventional hard-patchy colloids \cite{Bianchi-review}, the patches here are formed by the self-association of flexible polymer chains, which can recombine and rearrange. The process of self-assembling in patches is then succeeded by a hierarchical assembly on an even larger scale. 
The properties of TSP's are determined by the interactions between the monomers and the solvent, the percentage $\alpha$ of solvophobic monomers per arm, and the number of arms $f$.  
In what follows, each of the $f$ arms is composed by a total of $M_A$ solvophilic and $M_B$ solvophobic monomers.

\begin{figure}
  \includegraphics[width=1\columnwidth]{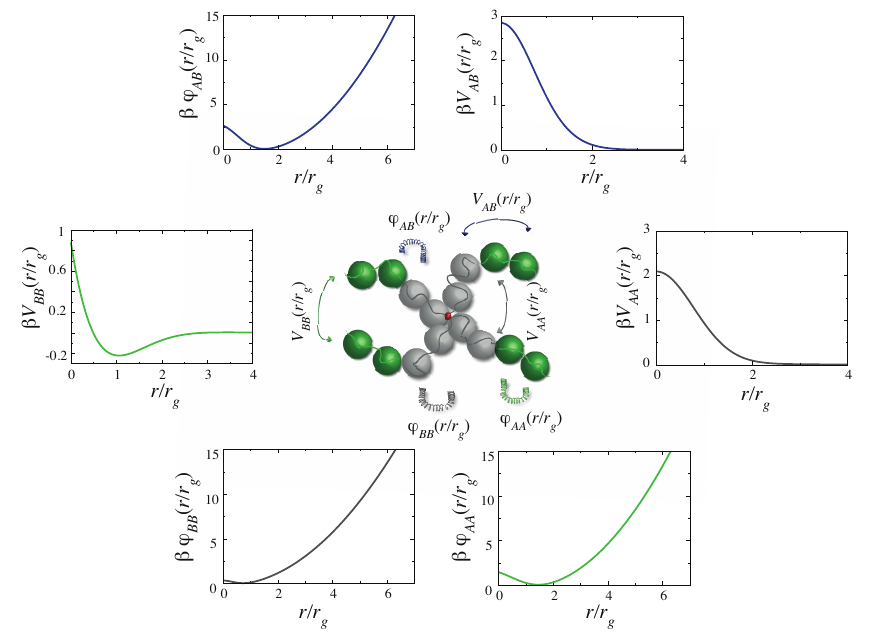} 
  \caption{Color online. The non-bonded potentials $V_{AA}$ in grey, $V_{AB}$ in blue  and $V_{BB}$ in green and the tethering potentials 
  $\varphi_{AA}$ in grey, $\varphi_{AB}$ in blue and $\varphi_{BB}$ in green between the various blobs of type $A$ and $B$. Color coding of the blobs: gray (solvophilic), green (solvophobic), anchor point (red). The radius of gyration of the $A$ and $B$ blobs is the same and  labeled as  $r_g$.} 
  \label{figure1} 
\end{figure}

By using a first-principles coarse-graining procedure based on a soft-effective segment picture~\cite{Capone_SoftMatter} 
(see Supplemental Material \cite{Note}), we represent the TSP's by modeling 
the $AB$ blocks of every arm with $n_A$ and $n_B$ blobs. Each individual blob (or effective segment) contains a fraction $m_A=M_A/n_A$ or $m_B=M_B/n_B$ monomers respectively. 
As was shown in Refs.~\cite{Capone_SoftMatter, Capone_Long}, once the set of potentials characteristic for a specific $AB$ diblock at a given temperature is computed, 
 any  asymmetry ratio $\alpha=M_A/(M_A + M_B)$ can be obtained by adjusting the numbers $n_A$ and $n_B$ of blobs in each arm.
Choosing the number of monomers in every blob large enough to be in the scaling regime enables us to obtain reliable properties and a general behavior of the polymeric particles, 
allowing thus for a mapping onto experimental systems. The underlying microscopic molecules analyzed here are formed by  $M=10\,000$ monomers per arm, 
which are represented in our approach by at least $n_{\rm min}=50$ blobs. Additional simulations of molecules with
$M=1000, 2000, 5000$ and $7000$ monomers per arm have produced identical results, 
therefore implying the scalability of the properties of the system with the size of the molecule, as long as each arm has sufficient monomers to be in the scaling regime. To extract the set of effective  potentials representing the effective segments within the arms, full monomer simulations of diblock copolymers were performed at zero density employing a lattice model.
In the full monomer representation, the  solvophilic part was represented by self avoiding random walks, while to represent the solvophobic part 
we used a square-well interaction between monomers, adjusting the well depth so that the $B$-blocks of the chains lie slightly below
their $\Theta$-temperature. The potentials, extracted taking into account all the many-body interactions acting in the system \cite{Capone_SoftMatter, Ladanyi}, are depicted in Fig.~\ref{figure1}.

\begin{figure}
  \includegraphics[width=1\columnwidth]{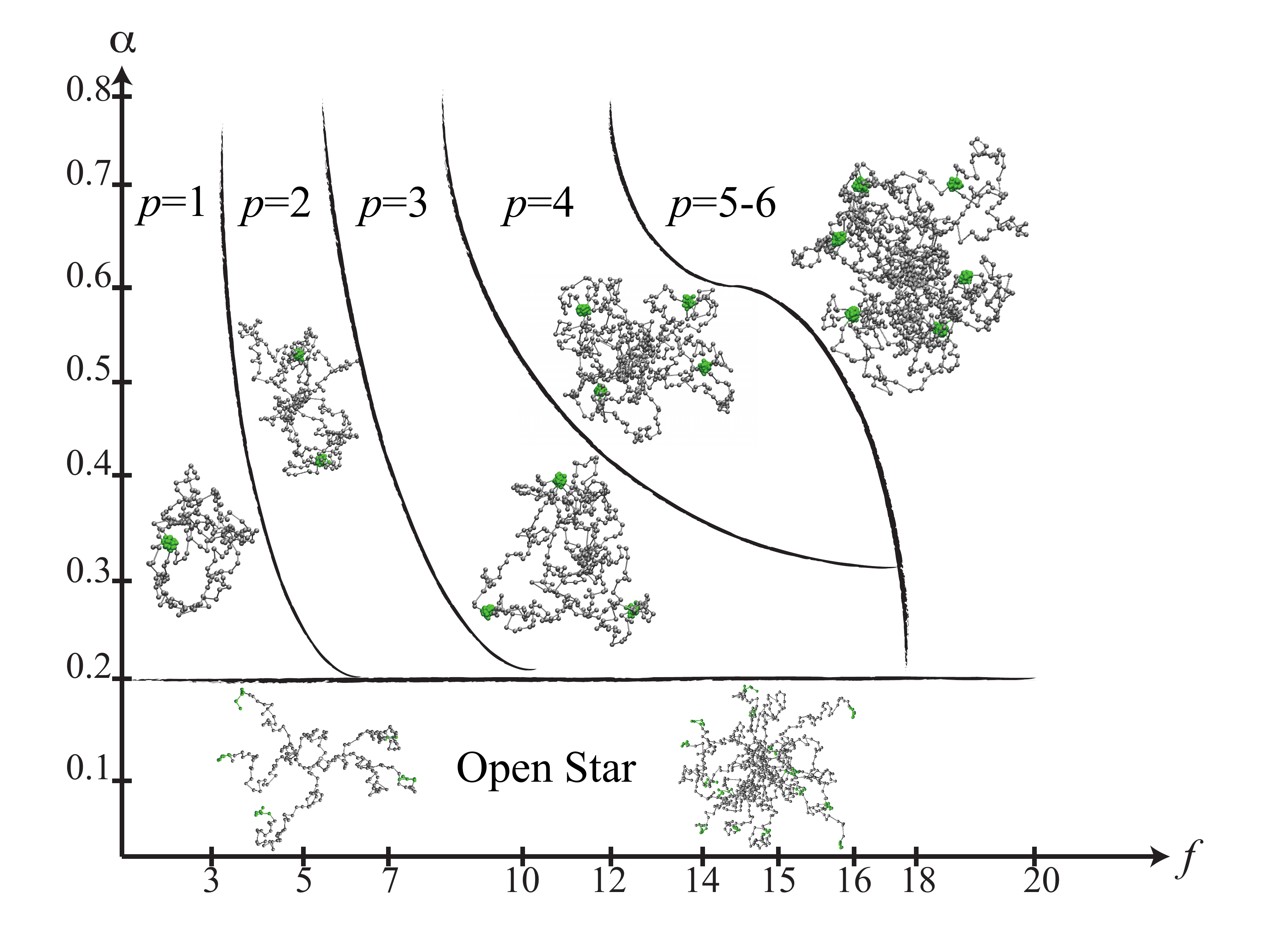}
  \caption{Single-molecule
  conformational state diagram of TSP's, shown as a function of the number of arms $f$ and asymmetry ratio $\alpha$. The number of multi-arm patches $p$ is indicated on the plot. Stars with $\alpha \lesssim 0.3$ remain in an open configuration.} 
  \label{figure2}
\end{figure}

We first characterize the system in the infinite dilution limit. To this end, the self-aggregation properties 
of a single molecule are classified as a function of its number of arms $f$ and the percentage $\alpha$ of attractive monomers per arm. 
Hereto, single molecule simulations are performed within the coarse-grained approach described above for various combinations of $f \in [2,20]$ and $\alpha\in [0.1, 0.8]$. The single-molecule conformational diagram showing the soft patchy particle nature of the TSP's and the tunability of the number of patches $p$ as a function of $f$ and $\alpha$ is presented in Fig.~\ref{figure2}.
It reveals that molecules with $\alpha \lesssim 0.3$  show no sign of self-aggregation, whereas such a behavior can clearly be identified for $\alpha \gtrsim 0.3$. 
For given $(f,\alpha)$ values, molecules self-assemble into patchy particles characterized by as many as six patches in the parameter range considered.
The number of patches that are formed and the concomitant anisotropy of the particles,
can be precisely tuned by choosing an appropriate combination of $(f,\alpha)$. 
Previous full monomer works on  TSP's with much shorter arms ($M \lesssim  30$) \cite{PRL-FG,PRE-FG} showed that, for a certain range of the $(f,\alpha)$ parameters, 
these molecules exhibit similar behavior as the stars with $M=10\,000$ monomers per arm: they  self-assemble  
into structures with  single or multiple assembly sites that were named single or multiple ``watermelons''.
The full monomer analysis showed  that   the single collapsed watermelon  conformation becomes unstable for short chains at fixed temperature when
increasing the number of arms  ($f>6$) in favor of multiple watermelon structures, identical to the patchy particles found here.  

Recently, it has been demonstrated that linear block copolymers have the ability to assemble into micelles or cylinders decorated by patches \cite{Cinesi-softmatter, Nanoletters2008}.  
In Ref.\ \cite{Nanoletters2008}, it was shown that an increase of  
the density of block copolymers in solution induces a growth in the patchy aggregate, whose size and patchiness therefore change 
with concentration. If one aims at producing robust soft patchy particles, however, the stability of the aggregate formed by block 
copolymers should be tested against the density of copolymers in solution. 
Therefore, to assess to which extent TSP's 
are stable, self-aggregated patchy particles that maintain their zero density properties even in the semi-dilute and concentrated regimes, 
simulations at finite density have been performed for a number of selected $(f,\alpha)$-combinations. 
We spaced our simulations over a wide range of densities, starting from $0.5\,\rho^*$ to $10\,\rho^*$, 
where $\rho^* = 3/(4\pi R_g^3)$ is the overlap density of TSP's having gyration radius $R_g$.
Simulations of $N \in [100,250]$ stars made of $f \in [3,20]$ arms that have a  percentage of attractive monomers $\alpha \in [0.2, 0.8]$ have been performed. 

\begin{figure}
  \includegraphics[width=0.9\columnwidth]{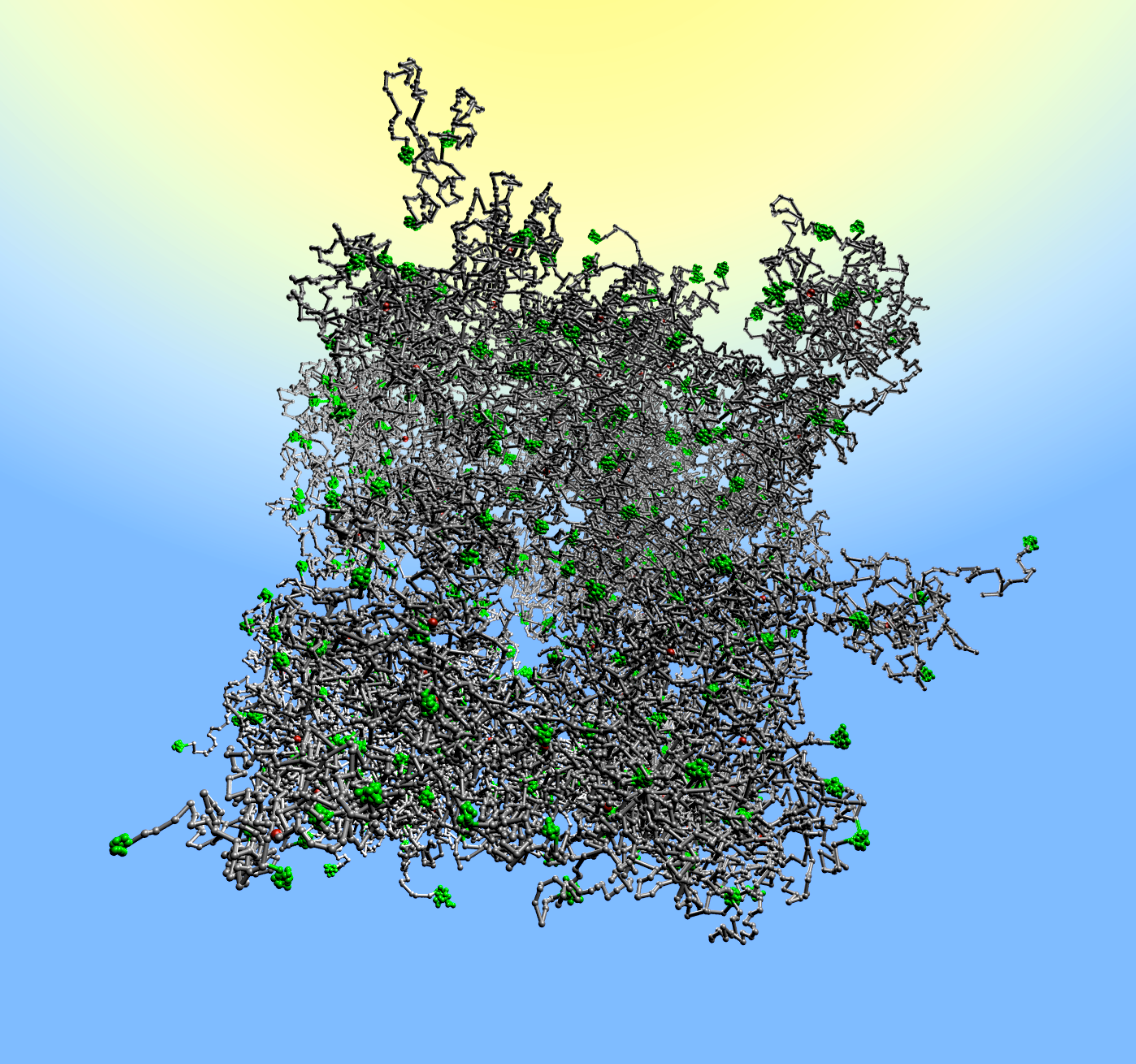}
  \caption{Snapshot of a percolating gel formed by telechelic star polymers with $f=5$ arms and $60\%$ of attractive monomers per arm. Color coding as in Fig.~\ref{figure1}. The snapshot shows the simulation box and some parts of the periodically repeated system.}
  \label{figure3} 
\end{figure}

For every density, we quantify the properties of the single particle and of the whole solution by performing both a cluster analysis and 
a topological analysis by means of the Euler characteristic. 
Both approaches (see the Supplemental Material
for details on the methodologies) 
consistently reveal that the finite density aggregation phenomenon is a consequence of the structure of the local intra-particle 
aggregation that leads to the formation of  patches.
As concentration grows, inter-star patch merging leads to the formation  of  low density percolating gels or networks, see Fig.~\ref{figure3}.
The cluster analysis  of the gel shows that the zero density single star properties, i.e., the average number of patches per star, the average angle between the patches, 
and the distance of the patches from the centre of the star, together with  the fluctuations over the average values, are unaffected by the density of stars  in solution. 
This is further supported by the topological analysis. Here, covering spheres of varying radius $R$ are placed at the center of each solvophobic blob
and the Euler characteristic of the resulting pattern, $\chi(R)$, is considered as a function of $R$. Typical results are shown in Fig.S2 in the Supplemental
Material.
The salient feature that can be seen there is the existence of broad, flat plateaus in the range $10 \lesssim R/r_g \lesssim 50$, which are the signature of the existence of well-formed and well-separated
patches. The number of patches per TSP, the number of blobs per patch as well as the average distance between the patches can all be extracted
from this topological invariant and the results show a fully consistent picture with that obtained at the infinite dilution limit.
Thus, even in the random gel, the zero-density self-aggregation properties of TSP's remain unaffected.

TSP's with a given patchiness thus emerge as  novel candidates for the stabilization of ordered, crystalline states possessing
local coordination compatible with the given particle anisotropy.
We therefore consider two target crystals to assess the large scale self-assembling properties of TSP soft patchy particles, namely a diamond and a simple cubic crystal.
Fourfold-coordinated TSP's
are compatible with a diamond lattice, whereas sixfold-coordinated stars are a natural choice to obtain stable simple cubic crystals.
Hard core patchy particles have proven to be an extremely versatile system for the synthesis of a wide variety of crystals; 
however even if several geometries of patches have been considered, some crystals have been found to be stable 
only within a very small interval of the parameters range, i.e., number of patches, strength of the attractive potential and temperature  \cite{Noya2010, romano:184501}.

We found that the flexibility both in the stiffness of the core and in the localization of the patches on the particle is a major advantage when patches undergo large-scale self-assembly processes.  This flexibility has its main origin at the entropically driven fluctuations of the self-avoiding blocks of the arms.
In particular, we placed the anchoring points of the TSP's at the positions of various perfect lattices (simple cubic, diamond, fcc, bcc) and their arms at various
different initial equilibrated configurations and we performed very long simulations, letting the system free to equilibrate and evolve.
Since the diamond lattice has a coordination of four neighbors, the natural choice for the $(f,\alpha)$ parameters of the constituent stars, are $f$ in the range $[10,16]$ and $\alpha \in [0.4,0.8]$, values for which, 
at zero density,  stars  self assemble in  four patches. Similarly, the simple cubic requires stars with $f=20$ and $\alpha \in [0.5,0.7]$, that self assemble in six patches at low density.  However, in principle, stars could stabilize different crystal structures whose coordination number is compatible with the number of arms per star instead of with the number of patches that are formed at zero density. If the stabilization mechanism is dominated by the attractive monomers at the end of every arm per star, for  example for the range of $f$ 
that assembles in four patches, then indeed the diamond, bcc and fcc crystals (with coordination numbers of 4, 8 and 12 respectively), could be all viable candidates. To reduce the computational cost, we considered the scenarios corresponding to $f=15$ and $\alpha=0.6$ and $f=15$ and $\alpha=0.4$ for the fourfold-coordinated TSP's, and $f=20, \alpha=0.5$ for the sixfold-coordinated ones at a wide range of reduced densities  $\rho/\rho^* \in [1.3, 10]$ .

For all the densities analysed, both  bcc and  fcc crystals melted within the simulation time; on the contrary, the diamond  and simple cubic crystals remain mechanically stable respectively for the four- and sixfold-coordinated particles. 
A mean square displacement (MSD) analysis of the crystals (not shown) has been performed to assess the stability of the system; 
both the bcc and fcc displayed a continuously growing MSD, while the diamond  and simple cubic crystals analysed show saturation to a plateau at long times. A cluster analysis performed on the crystal shows  that the zero density single-particle properties are preserved in the crystalline phase for a wide range of densities.
In particular, the average number of patches (patchiness, $p$), the bond angles between the segments connecting the anchoring
point to two different patches (inter-patch angle $\omega$) as well as the distance from the anchoring point to the center of
mass of the patch (patch extension $L$) were fully preserved in the crystalline phase for a wide range of densities with respect 
to the values they have at infinite dilution, see Tables ST1 and ST2 of the Supplemental
Material.

\begin{figure}
\centering
\subfigure[]{\includegraphics[width=0.45\columnwidth]{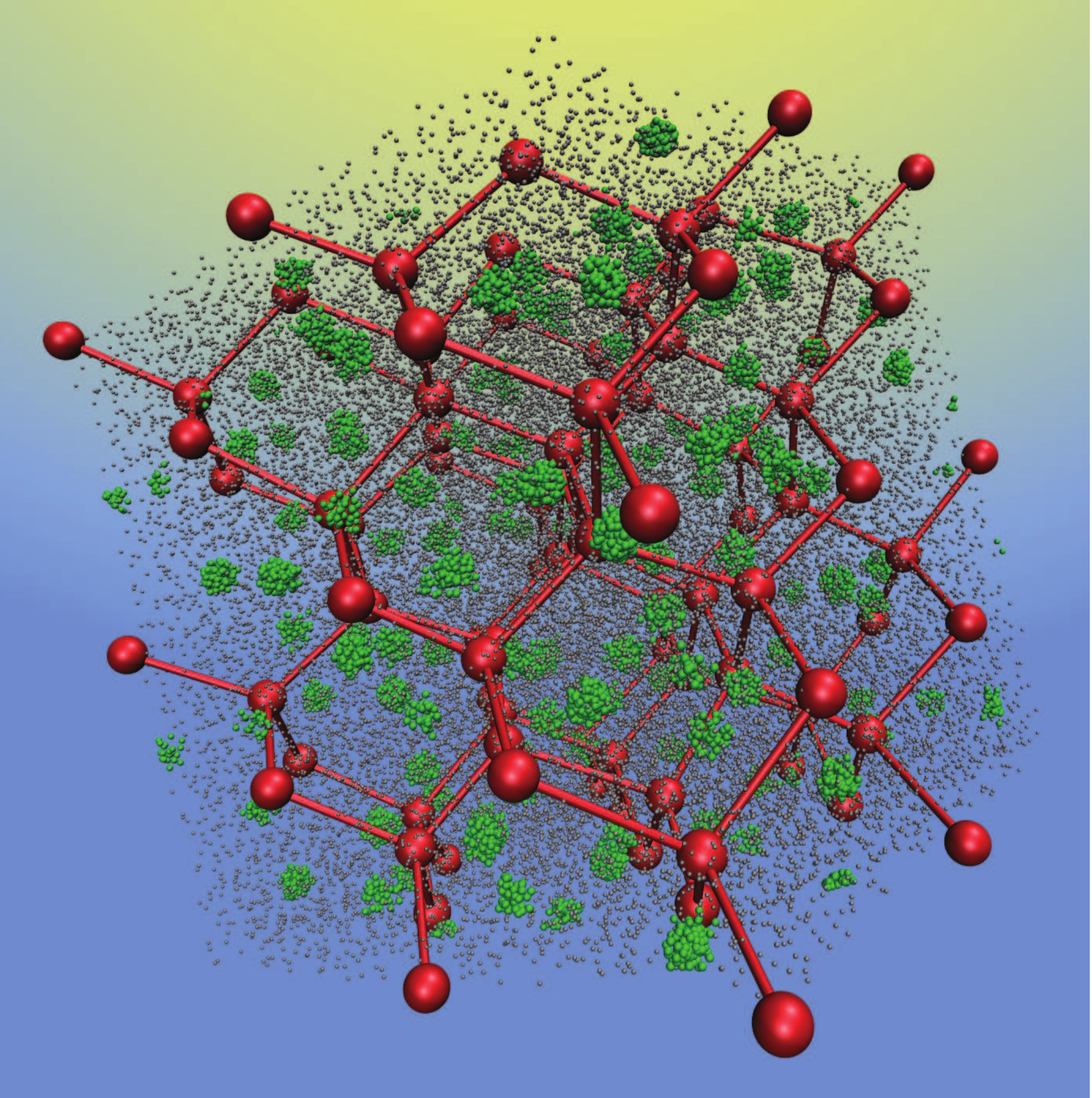}}
 \subfigure[]{ \includegraphics[width=0.45\columnwidth]{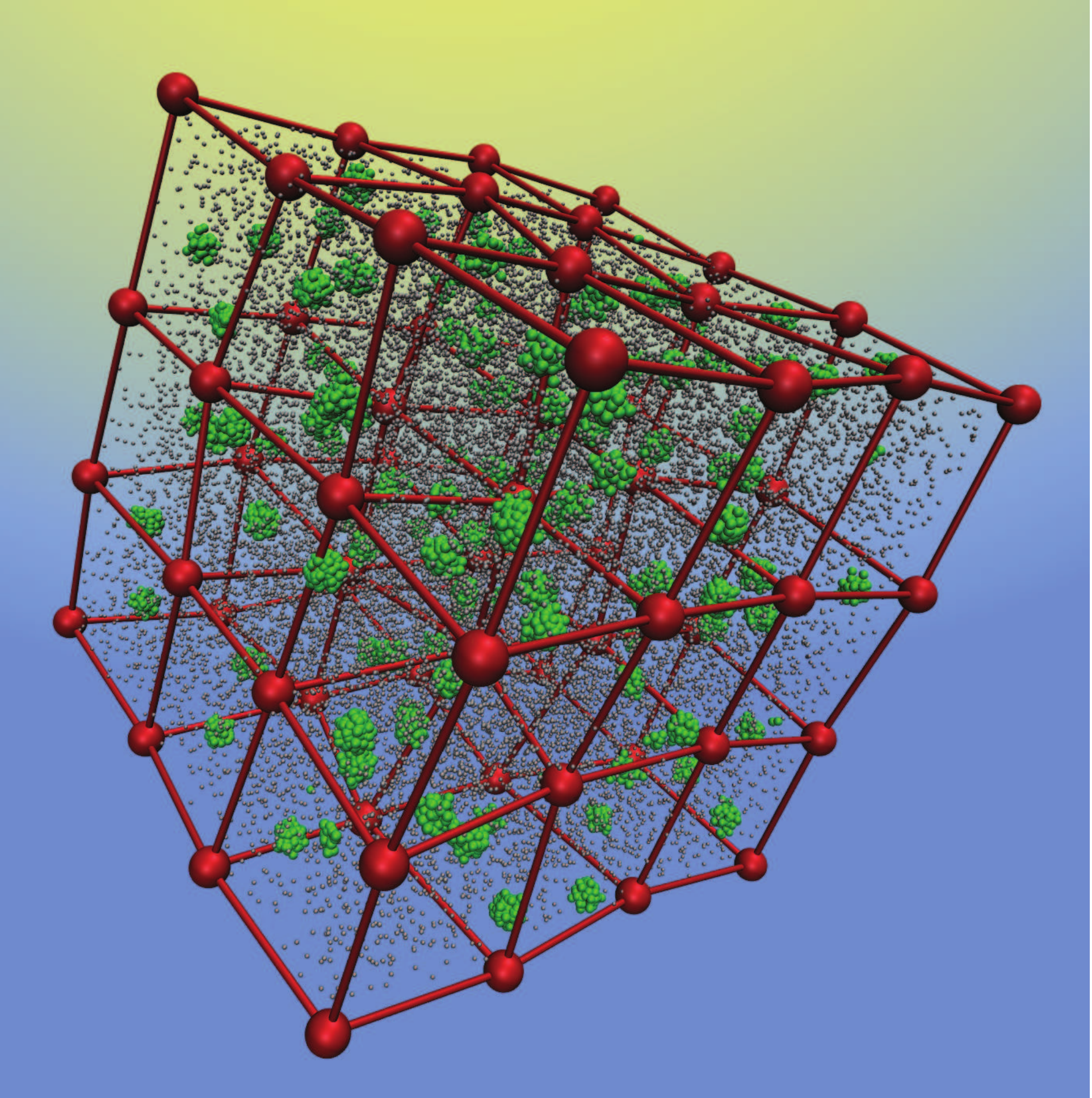}}
  \caption{\label{figure_diamond}Stable diamond lattice (a) and simple cubic crystal (b) obtained respectively for the $f=15, \alpha=0.4$  and $f=20, \alpha=0.5$ systems. The red lines connecting the anchor points of the stars, are a guide to the eye, drawn to stress the four fold coordinated  diamond lattice structure and six fold coordinated simple cubic structure. Color coding as in Fig.~\ref{figure1} (Bonds not shown).}
  \label{figure5}
\end{figure}

Tetrahedrally coordinated TSP's, such as the ones with $f=15$ considered here,
stabilize diamond crystals by merging the attractive patches belonging to molecules on neighboring sites, and sixfold-coordinated TSP's,
such as the ones with $f=20$ considered, stabilize simple cubical crystals via a similar mechanism. 
The flexibility in the location of the patches and in the distance between the patches and the anchor point induces a novel mechanism for the stabilization of the two crystals. 
Patches are not only merging among the first neighbors but involve further stars all the way to the third neighbors. Such a preference
can be understood by taking into account the fact that the TSP's maintain their patchy structures at all crystal densities, hence with decreasing lattice spacing the stars are forced to pack by pushing the bulky self-avoiding part of the arms into the empty spaces around each lattice site. 
For the diamond crystal, such spaces are distributed with tetrahedral symmetry, compatible with the natural angular arrangement of the patches  formed by each star. Once  
pushed into such spaces, the attractive patches can interact with the patches belonging to the other stars present in the same volume, which correspond up to the third neighbors. 
Similarly to the diamond, the simple cubic crystal is also stabilised by the packing of the self-avoiding branches of the TSP's, which reach patches in the neighboring cells.
Snapshots of a typical   diamond crystal formed for $(f,\alpha)=(15,0.4)$ and of a stable simple cubic crystal formed by $(f,\alpha)=(20,0.5)$ are shown in Fig.~\ref{figure5}.  
In the figure, the  diamond and the simple cubic arrangements of the centres of the stars, surrounded by a high density cloud of polymers,  are clearly visible.

The same mechanism is also responsible for the rapid melting of the bcc and fcc crystals which are not compatible neither with the packing of the self-avoiding arms
and lack the ability to create local environments that favor a tetrahedral arrangement.
This effect can be observed by monitoring the preference for first second and third neighbor patch fusion at increasing densities. Moreover,
and in contrast to the case of athermal star polymers and chains, the radius of gyration of the star is 
only very weakly affected by increasing the concentration. In fact, even at the infinite dilution the size of the multi-patch TSP's considered here is
significantly smaller than that or their athermal counterparts with the same $f$ and number of monomers per arm. Concomitantly, no further shrinkage
is necessary as the density grows and the TSP's maintain the size and form they have at the infinite dilution limit.

We have shown that telechelic star polymers are  extremely tunable and versatile building blocks that have the ability to hierarchically self-assemble into soft patchy particles, which organize themselves into complex structures. 
Using advanced computational methods we  fully characterize  the self-organization properties of  telechelic star polymers from zero density all the way up to high concentration regimes, where we observe stable  diamond and simple cubic crystals over a wide range of densities.
At low density the stars self-aggregate into particles decorated by a number of patches that is completely tunable by choosing different combinations  of the functionality $f$ of the stars and on the percentage $\alpha$ of attractive monomers per arm. Results remain invariant within the limits $M \in [10^3, 10^4]$ monomers per star, implying
their their scalability with molecular size. In addition, the major findings of our work rest on the competition between the enthalpic attractions between the solvophobic, terminal blocks and the self-avoiding central ones. Therefore, we expect that our results will remain essentially unaltered over a substantial range of temperatures $T<T_{\Theta}$, provided the terminal monomers do not reach the fully collapsed regime. 

The desired patchiness naturally arises
as the result of {\it self-organization} of the particles at the molecular level, rendering laborious preparation techniques unnecessary. 
These self-assembled patchy structures  survive at intermediate density regimes, where the particles form percolating networks.
Telechelic stars therefore have enormous potential as simple, extremely tunable and controllable soft and flexible patchy particles. 
Given their stability at finite density, such particles are excellent candidates as tunable building block for new materials and they provide
both {\it selectivity} in the crystal structures they stabilize and {\it scalability} in the lattice constant, since they maintain their properties within
the range of a few dozens to hundreds of thousands of monomers per arm.
 In contrast to hard patchy colloids, soft patchy particles take advantage of entropy in selecting and stabilizing structures. In particular,
allowing for fluctuations in the localization of the patches, results in an increase in the entropy of the target structures therefore enhancing their stability over a wide range of densities. 

State of the art of technology allows to fully control the synthesis of telechelic star polymers, and various experiments have been performed on solutions of telechelic star polymers at finite density \cite{Alward,Alward2} showing an extremely interesting aggregation scenario.
With this study we have established  telechelic star polymers as  a novel, promising, and extremely versatile self-assembled and tunable patchy particle system. 
The capability of these macromolecules to self-assemble into robust, soft patchy particles that maintain their shape and directionality
deep into the semidilute regime, surpasses problems related to the fabrication of conventional patchy colloids, opening-up ample 
new opportunities for experiments and technological applications. Moreover, their flexibility, which 
stems from their soft nature as fluctuating polymers, selectively stabilizes target crystals for a broad range of concentrations, opening the way
for the self-assembly of a variety of desired solids.

\section*{Acknowledgements}
The authors would like to thank Dimitris Vlassopoulos (Heraklion, Greece) for useful comments and suggestions and Emanuela Bianchi (Vienna, Austria) 
for useful discussions. This work was partially supported by the Marie Curie ITN-COMPLOIDS (Grant Agreement No.\ 234810).
Computer time at the Vienna Scientific Cluster (VSC) is gratefully acknowledged.

\newpage

\section{Supplemental Material for paper:
``Telechelic star polymers as self-assembling units from the molecular to the macroscopic scale''}

\setcounter{figure}{0}
\setcounter{table}{0}
\makeatletter 
\renewcommand{\thetable}{ST\@arabic\c@table} 
\renewcommand{\thefigure}{S\@arabic\c@figure} 
\makeatother

\subsection{Coarse-graining method}
The coarse graining used throughout this work, is an extension of the soft effective segment representation (SES) previously developed for both homopolymer linear (grafted and ungrafted) chains \cite{Pierleoni:2007p1560, Coluzza2011}  and block copolymers \cite{Capone_SoftMatter}. 
The methodology has been proven extremely efficient in reproducing  fundamental physical properties of polymers in the semi-dilute solution. 
The  strategy consists in representing the polymer as a chain of effective segments or blobs. 
Each blob, representing a group of monomers, interacts with other blobs via effective interactions that have been computed for dimers  (at infinite dilution) taking into account many-body contributions \cite{Capone_SoftMatter, Ladanyi}. 
The effective interaction computed at zero density successfully reproduces the physical properties of the system (e.g. pair distribution functions) till the blobs start to overlap, e.g., till the density $\rho_b$ of blobs in solution reaches the overlap density of blobs $\rho^*_b$. 
The SES strategy consists in representing the polymer chains with a number of blobs high enough to ensure that the condition  $\rho_b \leq \rho^*_b$ is satisfied. 
Hence, the higher the density in  solution the finer  is the level of the coarse graining required. 
In the range of densities considered in this paper  we found that  the minimum number $n_{\rm{min}}$ of blobs  that is required for obtaining a proper description in the scaling limit of telechelic star polymers, is determined by the local monomer density within the star. 
Therefore, to tune the level of coarse graining needed to reproduce the physical quantities in the system while guaranteeing that the coarse graining is not affecting the results,  we performed an extensive analysis of  single star properties, upon increasing the number of blobs per arm, therefore augmenting the detail of the coarse graining that we are using. 

\begin{figure}
\centering
  \includegraphics[width=0.9\columnwidth]{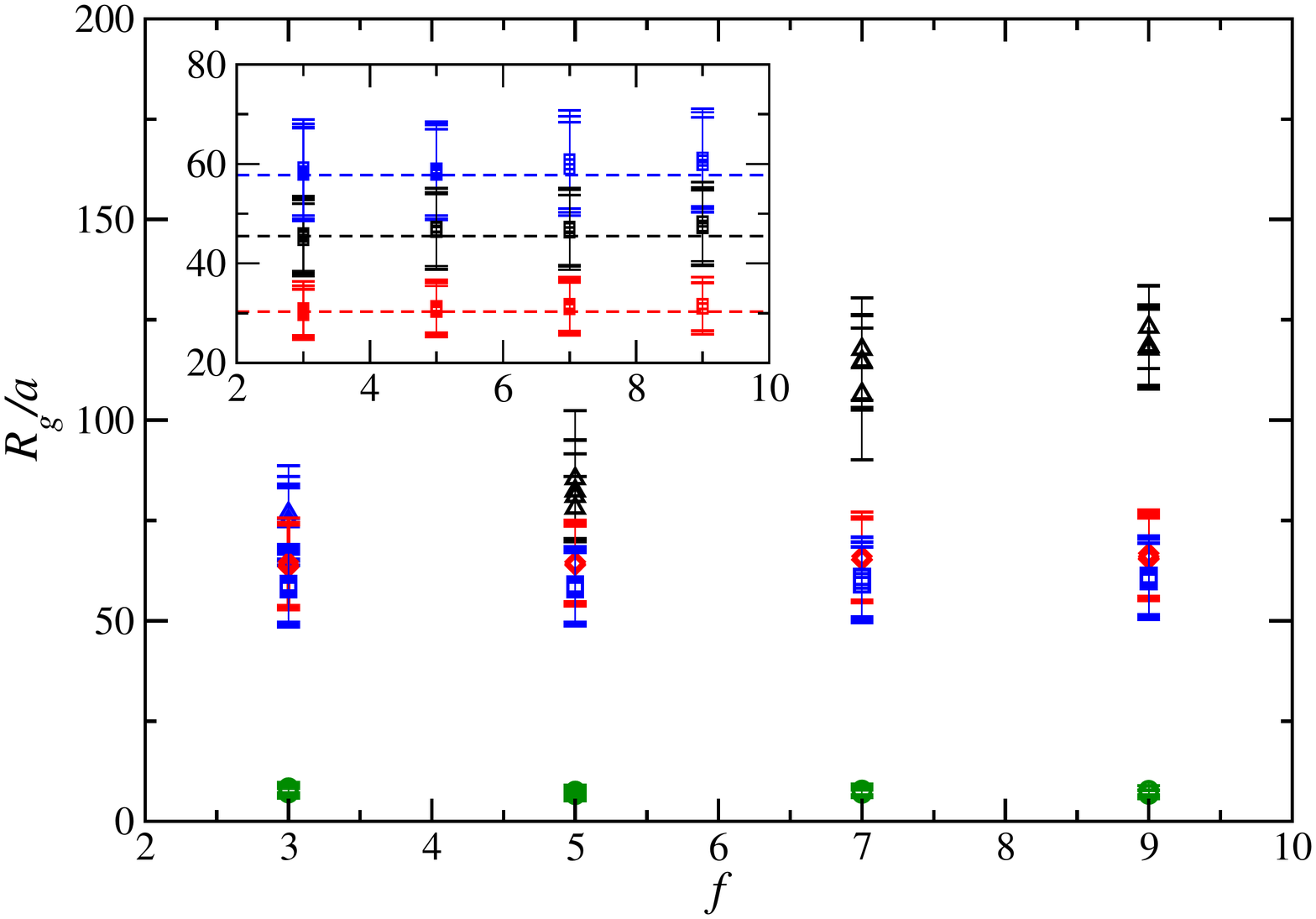}
  \caption{\label{figure6} Radius of gyration of whole star (black triangles) of each arm (red diamonds) of the self avoiding head of each arm (blue squares) and of the attractive tail of the arms (green circles) obtained using 50, 55, 60 and 65 blobs per arm for a star with $f\in[3,9]$ and $\alpha=0.4$. The various identical symbols correspond to different number of blobs. The inset shows the comparison between the radius of gyration of the self-avoiding part of the polymer arms (red $\alpha=0.8$, black $\alpha=0.6$, blue $\alpha=0.4$) and results obtained with full monomer simulations for self-avoiding homopolymers of equivalent length (dashed lines). The full monomers results have been rescaled in terms of the SES representation using the scaling law introduced in \cite{Coluzza2011}.}
\end{figure}

We found that for $n>n_{\rm {min}}$ all the properties of the single star (radius of gyration of the star, radius of gyration of the arms, radius of gyration of the two components of the arms, intra-aggregation properties), are independent of the level of the coarse graining used to represent the star, thus defining $n_{\rm {min}}$ as the minimum number of blobs needed for the results to be independent on the level of coarse graining used.
We show in Fig.~\ref{figure6} the estimate of the various radii of gyration calculated with four different level of coarse graining, namely 50, 55, 60 and 65 blobs per arm for stars with $f \in [3,9]$ and $\alpha=0.4$. 
As evident from the figure, the four different coarse graining levels perfectly agree in reproducing  the radius of gyration of the whole star, the radius of gyration of each arm of the star, of the self-avoiding head and of the attractive tail  of each arm. 
Therefore, as long as the number of blobs $n$ is bigger than the minimum number of blobs $n_{\rm{min}}$ (found to be  $n_{\rm min}=50$ for the cases at hand), all the properties of stars and system become coarse graining independent. Moreover, it has already been shown   for the case of grafted homopolymer brushes that the method of coarse graining allows for a controlled back-mapping onto full monomer results  \cite{Coluzza2011}. 
In  this work, we always used a total number of segments or blobs $n>n_{\rm{min}}$, thus satisfying the constraint for the validity of this approach. Performing simulations for systems slightly below the  $\Theta$-temperature for the solvophobic tails, and in the athermal solvent regime for the solvophilic heads, grants an attractive, effective potential between the $B$ and a repulsive, Gaussian potential between the $A$ blobs. The microscopical asymmetry ratio can also be rewritten in terms of the coarse grained model, which reads:
\begin{equation}\label{asymm}
  \alpha \simeq \frac{n_A \left({r_{gA}}\right)^{1/\nu_{A}} }{n_A \left({r_{gA}}\right)^{1/\nu_{A}} +n_B \left({r_{gB}}\right)^{1/\nu_{B}}},
\end{equation}
where $r_{gA}$ and $r_{gB}$  are the radii of gyration, and $\nu_{A}$ and 
$\nu_{B}$ are the Flory exponents of the $A$ and $B$ effective segments respectively. Here we chose $r_{gA}=r_{gB}=r_g$. Note that equation \eqref{asymm} assumes that the number of monomers in every segment is large enough to be in the scaling regime of polymers (e.g., large number of monomers per arm). 

\subsection{The Euler characteristic}\label{euler}

Systems assembling in low density percolating networks (or gels) are characterized by an inhomogeneous distribution of particles in space; such inhomogeneity makes gels  an extremely complex system to characterize, since high order correlations play an important role.  For this reason, a structural analysis of network forming systems are not straightforward:  common characterizations, essentially based on two particle correlations, do not give a full insight on the assembled status of the system and more complex analysis are required for such phases. 

\begin{figure}
  \includegraphics[width=0.9\columnwidth]{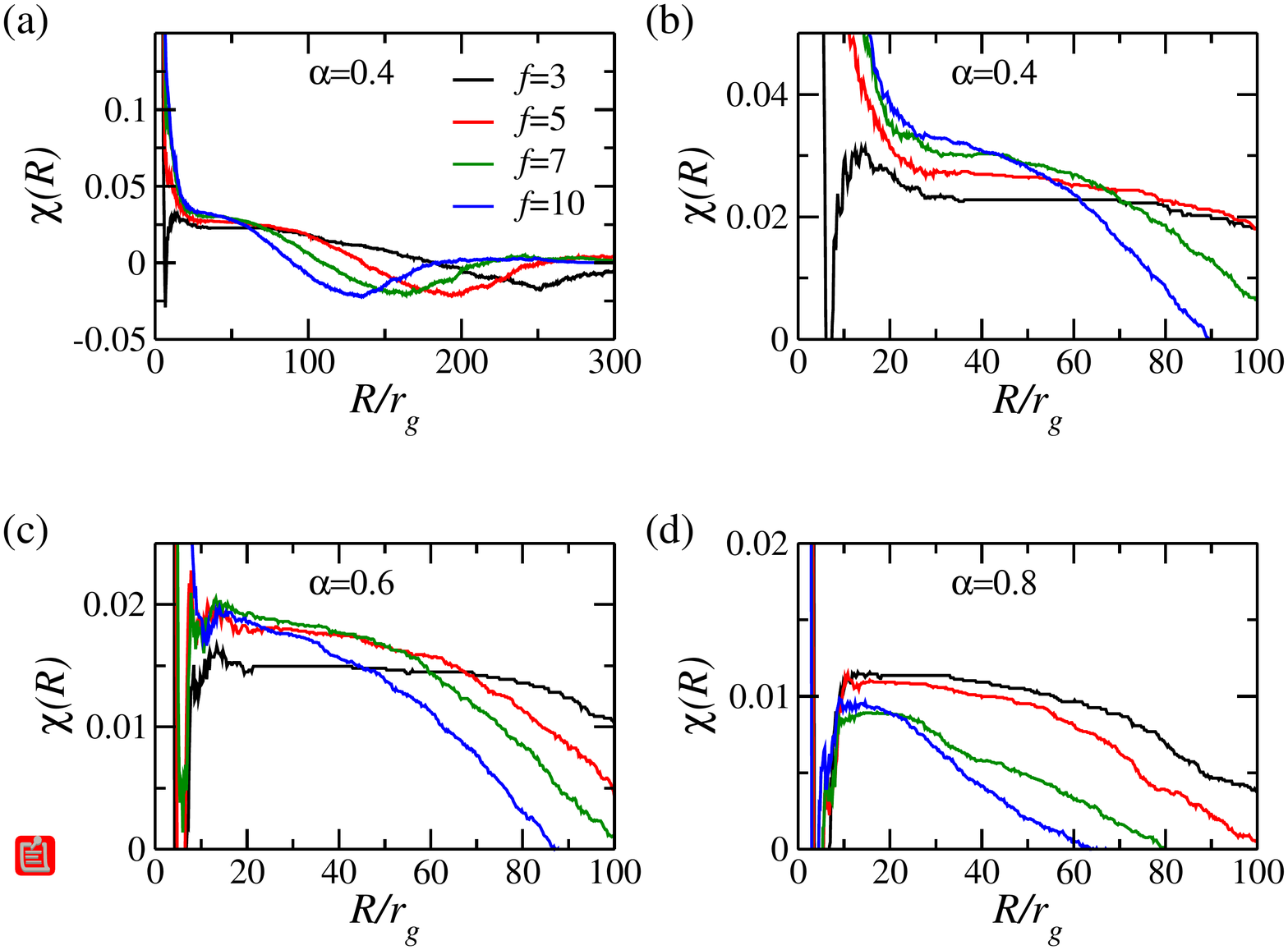}
  \caption{The Euler characteristic $\chi(R)$ for a system of telechelic star polymers with $f=3,5,7$ and $10$ arms and asymmetries $\alpha=0.4, 0.6$, and 0.8 at finite densities. The $x$ axis is scaled with the radius $r_g$ of gyration of the blob. Panel (b) is a zoom of panel (a) in the region of the plateaus of the Euler characteristic.}
  \label{figure4} 
 \end{figure}

A robust indicator that has been used to describe morphologies on vast scales, ranging from the large-scale structure of the universe~\cite{Mecke:AA1994} to the microscopic scale of molecules and atoms~\cite{likos:1995JCP, miller:2010JPCM} is based on the Euler-characteristic $\chi$. 
The latter is a topological invariant that was introduced by Leonhard Euler to characterize surfaces of polyhedra by counting their number of corners (vertices), edges and faces. 
For the topological characterization of the gels, we consider the surface that is formed by a collection of spheres with radius $R$ centered on the  position of each 
solvophobic blob in the TSP. The Euler characteristic  $\chi(R)$ that we obtain within our analysis is therefore an indicator for a family of surfaces formally parametrized by  $R$. The complexity of the analysed surfaces entails an enhanced sensitivity of this measurement with respect to many-particle correlations and provides insights in both  local and global structure of the gel phase. Examples of appropriately normalized $\chi(R)$ at finite densities, only considering the solvophobic blobs, are given in Fig.~\ref{figure4} of the main text for systems of telechelic star polymers with $f=3,5,7$ and $10$ arms and $\alpha =0.4, 0.6$ and $0.8$.

A change in the value of $\chi(R)$ is brought about by a topological difference in the surface induced by a small increase in the radius $R$ of the covering spheres. The simplest of the events causing those changes is found when two unconnected spheres start touching on increasing the value of $R$. In this particular case the topology is modified, because two, initially disjoint, surfaces have merged to a single one. In the same fashion, a plateau  in $\chi(R)$ indicates the absence of such events, which implies that at those length-scales ``nothing'' is happening, or that there is a balance between different types of events in the more clangorous range of distance scales. For a more complete description of the methodology we refer the reader to Refs.~\cite{miller:2009JCP,miller:2010JPCM}. 

The plateaus of $\chi(R)$ in Fig.~\ref{figure4} in the range  $10 \lesssim R/r_g \lesssim 50$ are indicative of the formation of well-separated patches by the telechelic stars. The beginning of each plateau is related to the distances between solvophobic blobs within the same patch, whereas the end and the height of a plateau correspond to the minimal distance between distinct patches and the average number of solvophobic blobs in a patch respectively.

\subsection{Cluster analysis} \label{cluster analysis}
The hierarchical self-assembling properties of TSP rely on the ability of the stars to keep their low-density assembled patchy particle structure upon augmenting densities. 
In the infinite dilution regime, the identification of the patches formed by a single star, together with the characterization of their angular and spatial distribution around the centre of the star, is straightforward. 
At finite densities on the other hand, particles start aggregating with one another and  the more they interact, the  more difficult it becomes to  distinguish the contribution of each star to the aggregates as well as to characterise the single star self-aggregating properties. 
The Euler characteristic is capable of identifying the  presence of clusters and their sizes, however  it does not allow for a proper characterization of the patchy nature of each star.

A simple visual analysis of simulations snapshots shows that the clusters formed by the solvophobic  arm ends are nicely defined and isolated, suggesting that a  cluster algorithm should be capable of precisely identifying the clusters. Fundamental properties are then measured in details, such as the cluster (patch)  size and arrangement around each stars, as well as the  counting and classification of the clusters resulting from the merging of distinct patches. The cluster analysis is then performed both at zero density and at finite density to fully characterize the properties both of the single star and of the solution. Clusters are defined by introducing a cut-off radius $R_{\rm cut}$: particles are classified as neighbours if their distance is below $R_{\rm cut}$. 

Clusters are then built by grouping particles that share at least one neighbour. According to this definition, a  patch is a cluster of solvophobic particles. 
$R_{\rm cut}$ is a free parameter, that is chosen in a range of values around $R_{\rm cut} \simeq r_g$, where $r_g$ is the radius of gyration of the blob. The optimal $R_{\rm cut}$ is such that  the number and size of the clusters is not affected by a small variation $\Delta R_{\rm cut}$ of the cut-off radius. Such a cut-off value exists because the clusters are well separated in solution as it can be seen in the simulations snapshots (see Fig.3 in the main article) and by the presence of plateau region in the Euler characteristic (see Fig.~\ref{figure4}). Once a list of cluster is obtained for each simulation snapshot, we measure the number of patches that belong to each star, their geometrical arrangement around the center of the star, and the degree of connectivity present in the system. Such quantities are then averaged over the configurations sampled for each density and star parameters (functionality $f$ and arm solvophobicity $\alpha$). Accordingly, we can classify how the patches distribute around the stars and how they are shared among the stars in solution. The averages are then compared to the zero density simulations as showed in Table~\ref{table1} and 
in Table~\ref{table2} below, verifying the robustness of the patchiness $p$ across a wide range of concentrations both in the ordered crystalline and in  the disordered gel phase.

\begin{table}[ht]
 \caption{\label{table1} Properties of the TSP's in the diamond lattice for two different tetrahedrally coordinated molecules, namely  $f=15, \alpha=0.6$ and $f=15, \alpha=0.4$ and for a six fold coordinated molecules, namely $f=20, \alpha=0.5$, over a wide range of densities,
 indicated at the first column. In the tables, $p$ represents the average number of patches, $\omega$ the bond angles between the segments connecting the
anchoring point to two different patches, $L$ the patch extension (the distance from the
anchoring point to the center of mass of the patch) and  at the last column, $a$ denotes the bond length in the underlying monomer resolved system.}

\vspace{4cm}

\begin{center}

\centering

\begin{tabular}{| c | c | c | c | c | c | c | c |}

\hline
  \multicolumn{4}{|c|}{ $f=15, \alpha=0.6$ } & \multicolumn{4}{|c|}{ $f=15, \alpha=0.4$ }  \\
\hline
 {$\rho/\rho*$}   &   $p$ &   $\omega$ [in degrees]  &  $L$ [in units of $a$] &  {$\rho/\rho*$}    &  $p$   &   $\omega$ [in degrees]  &    $L$ [in units of $a$]     \\
\hline

0.0    & 4.01 & 103.5 $\pm$ 27.8    & 142.2 $\pm$ 29.2 & 0.0    & 4.02  & 104.4 $\pm$ 31.8  &  161.6 $\pm$ 27.6 \\
1.3 & 3.85 & 102.6$\pm$ 24.8 & 135.4$\pm$ 37.6 & 1.9 & 4.20 & 100.7$\pm$ 37.9 & 165.0 $\pm$ 42.0 \\
2.2 & 3.84 & 102.6$\pm$ 34.8 & 132.2$\pm$ 35.5 & 3.0 & 4.25 &  101.8$\pm$ 38.0 & 164.9 $\pm$ 41.0  \\
3.8 & 3.81 & 103.5$\pm$ 33.5 & 132.8$\pm$ 34.5 & 5.2 & 4.12 &  102.7$\pm$ 35.2 &  168.9 $\pm$ 38  \\
7.4 & 3.86 & 103.5$\pm$ 34.3 & 134.8$\pm$ 32.2 & 10 & 4.13  &  101.8$\pm$ 36.3 & 168.1$\pm$ 34.2  \\

\hline
\end{tabular} \\

\begin{tabular}{| c | c | c | c |}
\hline

 \multicolumn{4}{|c|}{ $f=20, \alpha=0.5$ }  \\
 \hline
 {$\rho/\rho*$}    &  $p$   &   $\omega$ [in degrees]  &    $L$ [in units of $a$]    \\
\hline

0.0  &  6.1 & 96.5  $\pm$ 34.2 &  168.0 $\pm$ 32.3  \\
1.28 & 5.7  & 94.1  $\pm$ 38.4 &  171.5 $\pm$ 36.8  \\
3.05 & 5.6  & 93.4  $\pm$ 37.4 &  172.7 $\pm$ 35.7  \\
5.72 & 5.5  & 94.6  $\pm$ 38.7 &  169.4 $\pm$ 33.1  \\
10.3 & 5.6  & 88.4  $\pm$ 42.7 &  155.5 $\pm$ 27.1  \\
\hline

\end{tabular}

\end{center} 
\end{table}

\begin{table}[ht]
 \caption{\label{table2} Intra particle aggregation properties for star polymers  with $f\in[5,15]$ and  
 fixed asymmetry $\alpha=0.4$, over a wide range of densities (infinite dilution, $\rho/\rho^* =0$, and in the gel phase, $\rho/\rho^* > 0$).}

\begin{center}

\centering

\begin{tabular}{| c | c | c | c | c| c| c| c| }

\hline
\multicolumn{2}{|c|}{ $f=5$ }  & \multicolumn{2}{|c|}{ $f=7$ }   & \multicolumn{2}{|c|}{ $f=10$ } &   \multicolumn{2}{|c|}{ $f=15$ }\\
\hline

 {$\rho/\rho*$}   &   $p$ &   {$\rho/\rho*$}   &   $p$ &{$\rho/\rho*$}   &   $p$  &{$\rho/\rho*$}   &   $p$    \\
\hline

0      & 1.12 & 0      & 2.13 & 0      & 2.84  & 0     & 4.02  \\
0.10   & 1.24 & 0.25   & 1.93 & 0.38   &  3.16 & 0.52  & 3.90\\
0.20   & 1.20 & 0.50   & 1.82 & 0.75   & 2.65  &  1.03 & 3.84\\
0.25   & 1.13 & 0.63   & 1.80 & 0.93   & 2.61  &  1.30 & 3.88\\
0.30   & 1.30 & 0.76   & 1.86 & 1.12   & 2.41  &  1.55 & 3.86\\

\hline

\end{tabular}

\end{center} 
\end{table}


\begin{thebibliography}{10}
\expandafter\ifx\csname url\endcsname\relax
 \def\url#1{\texttt{#1}}\fi
\expandafter\ifx\csname urlprefix\endcsname\relax\def\urlprefix{URL }\fi
\providecommand{\bibinfo}[2]{#2}
\providecommand{\eprint}[2][]{\url{#2}}

\bibitem{Glotzer07}
\bibinfo{author}{ S.~C.~Glotzer}, and \bibinfo{author}{M.~J.~Solomon},
\newblock \emph{\bibinfo{journal}{Nature Mater.}} \textbf{\bibinfo{volume}{6}},
 \bibinfo{pages}{557} (\bibinfo{year}{2007}).

\bibitem{Daan}
\bibinfo{author}{S.~Angioletti-Uberti}, \bibinfo{author}{B.~M.~Mognetti}, and
 \bibinfo{author}{D.~Frenkel},
\newblock \emph{\bibinfo{journal}{Nature Mater.}}  (\bibinfo{year}{2012}).
\newblock \urlprefix\url{http://dx.doi.org/10.1038/nmat3314}.

\bibitem{Torquato}
\bibinfo{author}{S.~Torquato}, and \bibinfo{author}{Y.~ Jiao},
\newblock \emph{\bibinfo{journal}{Nature}} \textbf{\bibinfo{volume}{460}},
 \bibinfo{pages}{876} (\bibinfo{year}{2009}).

\bibitem{romano:184501}
\bibinfo{author}{F.~Romano}, \bibinfo{author}{E.~Sanz}, and
 \bibinfo{author}{F. Sciortino},
\newblock \emph{\bibinfo{journal}{J. Chem. Phys.}}
 \textbf{\bibinfo{volume}{132}}, \bibinfo{pages}{184501}
 (\bibinfo{year}{2010}).

\bibitem{Panagiotopoulos}
\bibinfo{author}{P.~Akcora}, et al \bibinfo{author}{et al},
\newblock \emph{\bibinfo{journal}{Nature Mater.}} \textbf{\bibinfo{volume}{8}},
 \bibinfo{pages}{354} (\bibinfo{year}{2009}).

\bibitem{Velegol1}
\bibinfo{author}{C.~E. Snyder}, \bibinfo{author}{A.~M. Yake},
 \bibinfo{author}{J.~D. Feick}, and \bibinfo{author}{ D. Velegol},
\newblock \emph{\bibinfo{journal}{Langmuir}} \textbf{\bibinfo{volume}{21}},
 \bibinfo{pages}{4813} (\bibinfo{year}{2005}).

\bibitem{Velegol2}
\bibinfo{author}{A.~M. Yake}, \bibinfo{author}{ C.~E. Snyder}, and
 \bibinfo{author}{D. Velegol},
\newblock \emph{\bibinfo{journal}{Langmuir}} \textbf{\bibinfo{volume}{23}},
 \bibinfo{pages}{9069} (\bibinfo{year}{2007}).

\bibitem{Nie}
\bibinfo{author}{Z. Nie}, \bibinfo{author}{W. Li}, \bibinfo{author}{M. Seo},
 \bibinfo{author}{ S. Xu}, and \bibinfo{author}{ E. Kumacheva},
\newblock \emph{\bibinfo{journal}{J. Am. Chem. Soc.}}
 \textbf{\bibinfo{volume}{128}}, \bibinfo{pages}{9408} (\bibinfo{year}{2006}).

\bibitem{Kretz_2008}
\bibinfo{author}{A.~B. Pawar}, and \bibinfo{author}{ I. Kretzschmar},
 \bibinfo{pages}{355} (\bibinfo{year}{2008}).

\bibitem{Kretz_2009}
\bibinfo{author}{A.~B. Pawar}, and \bibinfo{author}{I. Kretzschmar},
\newblock \emph{\bibinfo{journal}{Langmuir}} \textbf{\bibinfo{volume}{25}},
 \bibinfo{pages}{9057} (\bibinfo{year}{2009}).

\bibitem{PRL-FG}
\bibinfo{author}{F. Lo~Verso}, \bibinfo{author}{C.~N. Likos},
 \bibinfo{author}{C. Mayer}, and \bibinfo{author}{H. L\"owen},
\newblock \emph{\bibinfo{journal}{Phys. Rev. Lett.}}
 \textbf{\bibinfo{volume}{96}}, \bibinfo{pages}{187802}
 (\bibinfo{year}{2006}).

\bibitem{faraday}
\bibinfo{author}{F. Lo~Verso}, \bibinfo{author}{A.~Z. Panagiotopoulos}, and
 \bibinfo{author}{C.~N. Likos},
\newblock \emph{\bibinfo{journal}{Faraday Discuss.}}
 \textbf{\bibinfo{volume}{144}}, \bibinfo{pages}{143} (\bibinfo{year}{2010}).

\bibitem{Bianchi-review}
\bibinfo{author}{E. Bianchi}, \bibinfo{author}{R. Blaak}, and
 \bibinfo{author}{ C.~N. Likos},
\newblock \emph{\bibinfo{journal}{PCCP Phys. Chem. Ch. Ph.}}
 \textbf{\bibinfo{volume}{13}}, \bibinfo{pages}{6397} (\bibinfo{year}{2011}).

\bibitem{Capone_SoftMatter}
\bibinfo{author}{ B. Capone}, \bibinfo{author}{J.-P. Hansen}, and
 \bibinfo{author}{I. Coluzza},
\newblock \emph{\bibinfo{journal}{Soft Matter}} \textbf{\bibinfo{volume}{6}},
 \bibinfo{pages}{6075} (\bibinfo{year}{2010}).

\bibitem{Capone_Long}
\bibinfo{author}{B. Capone}, \bibinfo{author}{I. Coluzza}, and
 \bibinfo{author}{J.-P. Hansen},
\newblock \emph{\bibinfo{journal}{J. Phys.: Condens. Matter}}
 \textbf{\bibinfo{volume}{23}}, \bibinfo{pages}{194102}
 (\bibinfo{year}{2011}).

\bibitem{Ladanyi}
\bibinfo{author}{B.~M. Ladanyi }, and \bibinfo{author}{D. Chandler},
\newblock \emph{\bibinfo{journal}{J. Chem. Phys.}}
 \textbf{\bibinfo{volume}{62}}, \bibinfo{pages}{4308} (\bibinfo{year}{1975}).

\bibitem{PRE-FG}
\bibinfo{author}{F. Lo~Verso}, \bibinfo{author}{C.~N. Likos}, and
 \bibinfo{author}{H. L\"owen},
\newblock \emph{\bibinfo{journal}{J. Phys. Chem. C}}
 \textbf{\bibinfo{volume}{111}}, \bibinfo{pages}{15803}
 (\bibinfo{year}{2007}).

\bibitem{Cinesi-softmatter}
\bibinfo{author}{J. Zhang}, \bibinfo{author}{Z.-Y. Lu}, and
 \bibinfo{author}{S. Zhao-Yan},
\newblock \emph{\bibinfo{journal}{Soft Matter}} \textbf{\bibinfo{volume}{7}}, \bibinfo{pages}{9944}
 (\bibinfo{year}{2011}).

\bibitem{Nanoletters2008}
\bibinfo{author}{G. Srinivas}, and \bibinfo{author}{J.~W. Pitera},
\newblock \emph{\bibinfo{journal}{Nano Letters}} \textbf{\bibinfo{volume}{8}},
 \bibinfo{pages}{611} (\bibinfo{year}{2008}).

\bibitem{Noya2010}
\bibinfo{author}{E.~G. Noya}, \bibinfo{author}{C. Vega},
 \bibinfo{author}{ J. P.~K. Doye}, and \bibinfo{author}{A.~A. Louis},
\newblock \emph{\bibinfo{journal}{J. Chem. Phys.}}
 \textbf{\bibinfo{volume}{132}}, \bibinfo{pages}{234511}
 (\bibinfo{year}{2010}).

\bibitem{Alward}
\bibinfo{author}{D.~B. Alward}, \bibinfo{author}{ D.~J. Kinning},
 \bibinfo{author}{E.~L. Thomas}, and \bibinfo{author}{L.~J. Fetters},
\newblock \emph{\bibinfo{journal}{Macromolecules}}
 \textbf{\bibinfo{volume}{19}}, \bibinfo{pages}{215} (\bibinfo{year}{1986}).

\bibitem{Alward2}
\bibinfo{author}{E. L. Thomas}, \bibinfo{author}{D. B. Alward}, \bibinfo{author}{D. J. Kinning}, \bibinfo{author}{D. C. Martin}, \bibinfo{author}{D. L. Handlin Jr.}, and \bibinfo{author}{L. J. Fetters},
\newblock \emph{\bibinfo{journal}{Macromolecules}}
 \textbf{\bibinfo{volume}{19}}, \bibinfo{pages}{2197} (\bibinfo{year}{1986}).

\bibitem{Pierleoni:2007p1560}
\bibinfo{author}{C. Pierleoni}, \bibinfo{author}{B. Capone}, and
 \bibinfo{author}{ J.-P. Hansen},
\newblock \emph{\bibinfo{journal}{J. Chem. Phys.}}
 \textbf{\bibinfo{volume}{127}}, \bibinfo{pages}{171102}
 (\bibinfo{year}{2007}).

\bibitem{Coluzza2011}
\bibinfo{author}{I. Coluzza}, \bibinfo{author}{B. Capone},
and \bibinfo{author}{J.-P. Hansen}, 
\newblock \emph{\bibinfo{journal}{Soft Matter}} \textbf{\bibinfo{volume}{7}},
 \bibinfo{pages}{5255} (\bibinfo{year}{2011}).

\bibitem{Mecke:AA1994}
\bibinfo{author}{K.~R. Mecke}, \bibinfo{author}{T. Buchert}, and
 \bibinfo{author}{H. Wagner}, 
\newblock \emph{\bibinfo{journal}{Astron. Astrophys.}}
 \textbf{\bibinfo{volume}{288}}, \bibinfo{pages}{697}
 (\bibinfo{year}{1994}).

\bibitem{likos:1995JCP}
\bibinfo{author}{C.~N. Likos}, \bibinfo{author}{ K.~R. Mecke}, and
 \bibinfo{author}{H. Wagner}, 
\newblock \emph{\bibinfo{journal}{J. Chem. Phys.}}
 \textbf{\bibinfo{volume}{102}}, \bibinfo{pages}{9350}
 (\bibinfo{year}{1995}).

\bibitem{miller:2010JPCM}
\bibinfo{author}{M.~A. Miller}, \bibinfo{author}{R. Blaak}, and
 \bibinfo{author}{J.-P. Hansen},
\newblock \emph{\bibinfo{journal}{J. Phys.: Condens. Matter}}
 \textbf{\bibinfo{volume}{22}}, \bibinfo{pages}{104109}
 (\bibinfo{year}{2010}).

\bibitem{miller:2009JCP}
\bibinfo{author}{M.~A. Miller}, \bibinfo{author}{R. Blaak},
 \bibinfo{author}{C.~N. Lumb}, and  \bibinfo{author}{ J.-P. Hansen}, 
\newblock \emph{\bibinfo{journal}{J. Chem. Phys.}}
 \textbf{\bibinfo{volume}{130}}, \bibinfo{pages}{114507}
 (\bibinfo{year}{2009}).


\bibitem{Note}
\bibinfo{title}{Note, see Supplemental Material at TO BE FILLED BY APS for the details on the coarse graining techniques and on the data analysis methodology}
 
 
\end{thebibliography}

\end{document}